# Self-Supervised Pre-Training for Deep Image Prior-Based Robust PET Image Denoising

Yuya Onishi, Fumio Hashimoto, Kibo Ote, Keisuke Matsubara, Masanobu Ibaraki

*Abstract*—**Deep image prior (DIP) has been successfully applied to positron emission tomography (PET) image restoration, enabling represent implicit prior using only convolutional neural network architecture without training dataset, whereas the general supervised approach requires massive low- and high-quality PET image pairs. To answer the increased need for PET imaging with DIP, it is indispensable to improve the performance of the underlying DIP itself. Here, we propose a self-supervised pre-training model to improve the DIP-based PET image denoising performance. Our proposed pre-training model acquires transferable and generalizable visual representations from only unlabeled PET images by restoring various degraded PET images in a self-supervised approach. We evaluated the proposed method using clinical brain PET data with various radioactive tracers ($^{18}$F-florbetapir, $^{11}$C-Pittsburgh compound-B, $^{18}$F-fluoro-2-deoxy-D-glucose, and $^{15}$O-CO$_2$) acquired from different PET scanners. The proposed method using the self-supervised pre-training model achieved robust and state-of-the-art denoising performance while retaining spatial details and quantification accuracy compared to other unsupervised methods and pre-training model. These results highlight the potential that the proposed method is particularly effective against rare diseases and probes and helps reduce the scan time or the radiotracer dose without affecting the patients.**

*Index Terms*—**Deep image prior, PET image denoising, pre-training, self-supervised learning, unsupervised learning.**

## I. INTRODUCTION

POSITRON emission tomography (PET) is an advanced functional imaging modality that observes the molecular-level activity in tissues caused by radioactive tracers [1]. Among them, dedicated brain PET scanners play an increasingly important role in clinical and research applications for more accurate diagnosis of cancer or dementia [2]–[5]. PET images typically have a low signal-to-noise ratio owing to physical degradation factors and limited counts compared to other biomedical imaging methods. Short-time scans or low-dose radiotracers that reduce the patient burden accelerate the degradation of PET images, potentially affecting diagnostic accuracy. This remains a major challenge, and an effective denoising method for low-quality PET images is essential.

The use of deep learning for PET image restoration has developed rapidly over the past several years [6]. The general approach is to map low-quality PET images to high-quality PET images through supervised training using a convolutional neural network (CNN) [7]–[9] or a transformer model [10], [11]. Combining multi-modal anatomical information, such as computed tomography (CT) or magnetic resonance (MR) images, which are easy to acquire simultaneously, achieves a higher denoising performance [12]–[14]. These supervised approaches require large datasets, including low- and high-quality image pairs. Acquiring massive amounts of high-quality PET images for training data without burdening the patient is particularly difficult in clinical practice. Moreover, the generalization performance for different PET probes is poor, and denoised images may suffer inherent biases toward unknown cases not included in the training dataset.

Recently, unsupervised learning approaches that do not require label data have attracted considerable attention. In particular, the deep image prior (DIP) [15], which uses the CNN structure itself as an intrinsic regularizer and does not require the preparation of a prior training dataset, has been applied to PET image denoising and has demonstrated superior performance [16]–[21]. Furthermore, applications in image reconstruction and parametric imaging are being investigated by incorporating existing physical models and mathematical algorithms into DIP [22]–[27]. While these have enormous potential for expansion to various applications in next-generation PET technology, their performance is highly dependent on the performance of the underlying network for the DIP. Therefore, we need to further improve the DIP-based restoration performance while maintaining the high robustness inherent in DIP.

In this study, we introduce a pre-training model created by self-supervised learning for DIP-based unsupervised PET image denoising. Using a pre-training model is considered one of the simple strategies to improve the performance of DIP that does not require reference data. Self-supervised learning can train using self-labels automatically generated from the unannotated data itself, and it is expected to acquire semantic feature representations that can be useful for downstream tasks [28]–[30]. Our proposed self-supervised pre-training model

Y. Onishi, F. Hashimoto, and K. Ote are with the Central Research Laboratory, Hamamatsu Photonics K.K., Hamamatsu, 434-8601, Japan (e-mails: yuya.onishi@hpk.co.jp; fumio.hashimoto@crl.hpk.co.jp; kibou@hpk.co.jp).

K. Matsubara is with the Department of Management Science and Engineering, Faculty of System science and Technology, Akita Prefectural University, Akita, 015-0055, Japan (e-mail: matsubara@akita-pu.ac.jp).

K. Matsubara and M. Ibaraki are with the Department of Radiology and Nuclear Medicine, Research Institute for Brain and Blood Vessels, Akita Cerebrospinal and Cardiovascular Center, Akita, 010-0874, Japan (e-mails: matsubara@akita-pu.ac.jp; iba@akita-noken.jp).

**a**. Pretext task (Self-supervised PET image restoration)

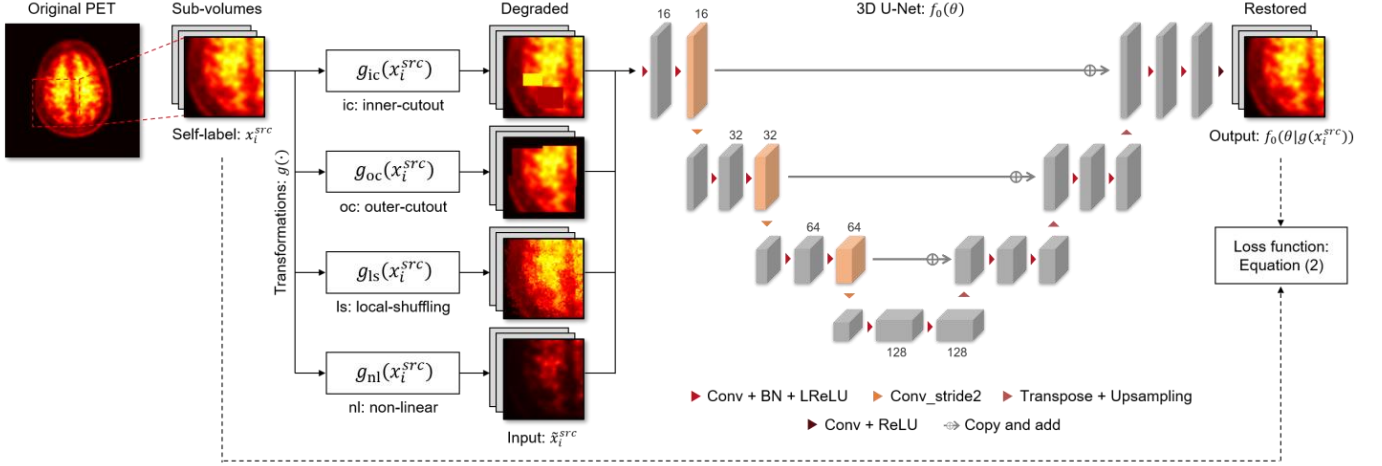

**b**. Downstream task (DIP-based unsupervised PET image denoising)

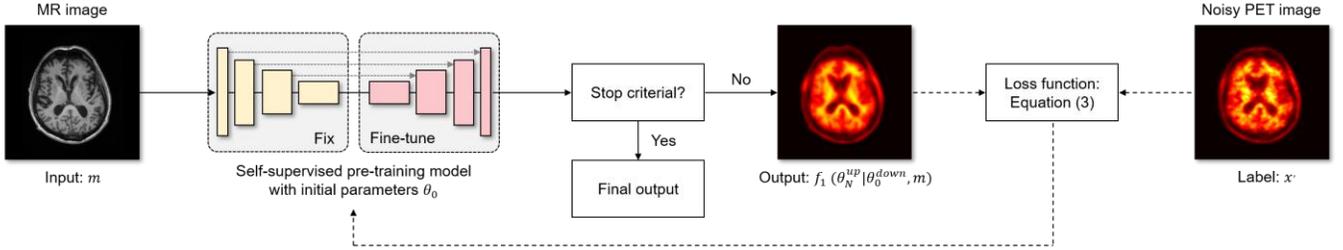

Fig. 1. An overview of the proposed method that consists of two steps: (a) pretext task of the self-supervised learning for the creation of the pre-training model and (b) downstream task of the unsupervised learning for PET image denoising. The network is first pre-trained to restore the degraded PET images to original ones based on the self-supervised learning framework, and then main DIP-based PET image denoising is performed by introducing parameters transferred from the pre-trained network.

aims to acquire generalizable and transferable visual representations from only unlabeled PET images, eliminating the need for the costly collection of high-quality PET images, similar to denoising by DIP. Thus, the DIP performance can be easily improved using only readily available data that do not have low- and high-quality image pairs like public datasets. Herein, we realize a robust and state-of-the-art PET image denoising that does not require reference data in the whole process from the creation of the pre-training model to the downstream denoising task.

## II. METHODOLOGY

Fig. 1 shows an overview of the proposed method, which consists of two steps: 1) self-supervised learning for the creation of the pre-training model; and 2) unsupervised learning for PET image denoising. The network is first pre-trained to restore the degraded PET images to the original ones based on a self-supervised learning framework, and then main DIP-based PET image denoising is performed by introducing parameters transferred from the pre-trained network.

### A. Self-Supervised Pre-Training

The objective of the self-supervised pre-training model is to acquire a common image representation which is generalizable and transferable across cases, such as different PET probes, diseases, and scanners. The proposed framework, shown in Fig. 1(a) performs three-dimensional (3D) image restoration as a pretext task using unlabeled original PET images.

First, we randomly crop sub-volumes $x^{src} = \{x_1^{src}, x_2^{src}, \cdots, x_n^{src}\}$ with a size of $64 \times 64 \times 32$ voxels containing the brain region as 3D patches from a prior training dataset and then apply a transformation function $g(\cdot)$:

$$\tilde{x}^{src} = g(x^{src}), \qquad (1)$$

where $\tilde{x}^{src} = \{\tilde{x}_1^{src}, \tilde{x}_2^{src}, \cdots, \tilde{x}_n^{src}\}$ is the degraded sub-volumes. As a transformation function, we consolidate four degradation schemes: inner-cutout, outer-cutout, local-shuffling, and non-linear transformation reported in previous studies [28]–[30], into a single image restoration task, allowing the model to learn robust PET image representations by restoring different degraded sets. Cut-out processes apply a single window with a complex shape to a sub-volume by generating and overlapping arbitrary windows of different sizes and aspect ratios. The inner-cutout masks the inner window regions with a random number, and the outer-cutout masks the surroundings. In local-shuffling, we shuffle the pixels inside each window to increase local variations in the PET images without compromising their global structures. Non-linear transformation uses the Bézier curve, an intensity transformation function randomly generated by from two endpoints and two control points, to assign new distinct values to different pixels [31].

Subsequently, the 3D encoder-decoder network $f_\theta$ is pre-trained to restore the degraded sub-volumes to their original ones based on the self-supervised learning framework as follows:

$$\theta_0 = \underset{\theta}{\text{argmin}} \frac{1}{N_t} \sum_{i \in D_t} \|x_i^{src} - f_0(\theta|\tilde{x}_i^{src})\|, \qquad (2)$$

where $\|\cdot\|$ is the L2 loss, $\theta$ denotes the trainable parameters, $D_t$ is a mini-batch sample of size $N_t$. In this method, the number of epochs and the batch size were set to 1,000 and 32, respectively. The stochastic gradient descent algorithm (learning rate $= 0.1$, momentum $= 0.9$) was implemented to solve the loss function in (2). Finally, we obtain the pre-trained network parameters $\theta_0$.

### B. DIP-Based PET Image Denoising

The pre-trained parameters $\theta_0$ containing robust PET image representations are introduced as the initial parameter of a 3D encoder-decoder network $f_1$ for DIP-based PET image denoising. Although the optimization algorithm of DIP remains basically the same, we fixed the parameters $\theta_0^{down}$ in the encoding paths and then fine-tuned the parameters $\theta^{up}$ in decoding paths as follows [32], [33]:

$$\hat{\theta}^{up} = \underset{\hat{\theta}^{up}}{\text{argmin}} \|x' - f_1(\theta^{up}|\theta_0^{down}, m)\|,$$
$$\hat{x}' = f_1(\hat{\theta}^{up}|\theta_0^{down}, m), \qquad (3)$$

where $\|\cdot\|$ is the L2 loss, the training label $x'$ and the network input $m$ are the noisy PET image and the co-registered MR image, respectively. Note that DIP input the entire 3D image and not patch images. The limited-memory BFGS (L-BFGS) algorithm [34], which is a quasi-Newton method that uses a second-order gradient and converges faster than other first-order optimizers, was implemented to solve the loss function in (3). After reaching the stop criterion, the network output is the final denoised PET image $\hat{x}'$.

### C. Network Architecture

The 3D encoder-decoder networks $f_0$ and $f_1$ share the same network architecture. It is based on a 3D U-Net and consist of encoding and decoding paths [Fig. 1(a)]. In the encoding path, the components of a $3 \times 3 \times 3$ convolution layer with batch normalization (BN) and a leaky rectified linear unit (LReLU) is repeated twice before being constructed by a $3 \times 3 \times 3$ downsampling convolution layer with two strides. For each downsampling step, the size of the feature maps is doubled, and the number of rows, columns, and slices decreased by half. In the decoding path, the outputs of a $3 \times 3 \times 3$ transpose convolution layer with trilinear upsampling and a skip connection supplied from the encoding path are added before the component of a $3 \times 3 \times 3$ convolution layer with BN and LReLU repeated twice. For each upsampling step, the size of the feature maps decreased by half and the number of rows, columns, and slices doubled. Eventually, a $1 \times 1 \times 1$ convolution layer with a rectified linear unit was used to output the denoised PET image. The architecture was built using PyTorch 1.6.0, Ubuntu 20.04 LTS and an NVIDIA TITAN RTX graphics processing unit with 24 GB memory.

### III. EXPERIMENTAL SETUP

We evaluated the proposed method using clinical brain PET data with various radioactive tracers, $^{18}$F-florbetapir (AV-45), $^{11}$C-Pittsburgh compound-B (PIB), $^{18}$F-fluoro-2-deoxy-D-glucose (FDG), and $^{15}$O-CO$_2$ acquired from different PET scanners. Note that our ultimate goal is not the task of image restoration using self-supervised learning. The usefulness of the pre-training model must be assessed based on its transferability and generalizability to various DIP-based PET image-denoising tasks. The denoising performance of the proposed method was compared with that of other unsupervised methods using anatomical information under the same conditions.

*Image Guided Filtering (IGF):* The IGF performs as an edge-preserving smoothing operator by adapting a local linear model using a guidance image [35]. We used an MR image as the guidance image.

*DIP w/ Scratch:* This is a general DIP algorithm that uses CNN parameters that are initialized to random values. We used DIP conditioned on MR images as the network input for better performance [17], [20].

*DIP w/ MR2PET:* This is a previously proposed pre-training model for DIP-based PET image denoising. An individual PET image was denoised in the conditional DIP framework based on a pre-trained basis function from the population information for domain transformation from MR to noisy PET images [33].

### A. Dataset

For the evaluation using $^{18}$F-AV-45, $^{11}$C-PIB, and $^{18}$F-FDG, we used cognitively normal subjects from the Open Access Series of Imaging Studies (OASIS-3), a public dataset containing multi-modal information from 42 to 95 years of age [36]. Table 1 summarizes the evaluation datasets. For the $^{18}$F-AV-45 scan, the participants received an intravenous bolus of approximately 10 mCi, and a dynamic 70 min emission scan was performed using a Siemens Biograph mMR. PET images were reconstructed using a 3D ordinary Poisson ordered-subset expectation-maximization (OP-OSEM) algorithm and cropped to $200 \times 200 \times 104$ voxels with a voxel size of $1.4 \times 1.4 \times 2.0$ mm$^3$ to reduce the demand on the GPU memory. Attenuation correction was performed by using a $\mu$map from Dixon-based tissue segmented MR scan. Participants using $^{11}$C-PIB and $^{18}$F-FDG received 6 mCi to 20 mCi and approximately 5 mCi, respectively, followed by a dynamic 60 min emission scan. The $^{11}$C-PIB scan was performed using Siemens Biograph 40 and reconstructed using a discrete inverse Fourier transform (DIFT) after a spiral CT scan for attenuation correction. The $^{18}$F-FDG scan was conducted using a Siemens ECAT HR+ 962 and reconstructed using filtered back projection (FBP) after the transmission scan. Each reconstructed image was cropped to $128 \times 128 \times 64$ voxels with a voxel size of $2.3 \times 2.3 \times 2.0$ mm$^3$ for $^{11}$C-PIB and $2.0 \times 2.0 \times 2.4$ mm$^3$ for $^{18}$F-FDG. Noisy PET images were obtained by extracting only a single 5 min frame (three frames for $^{11}$C-PIB) that clearly observed the contrast between the gray and white matter. The corresponding T1-weighted MR images acquired individually were interpolated

| | Pretext task | | Target task | | |
|---|---|---|---|---|---|
| PET data | [18]F-AV-45 [36] | [18]F-AV-45 [36] | [11]C-PIB [36] | [18]F-FDG [36] | [15]O-CO$_2$ [37] |
| Subject | Cognitively normal | Cognitively normal | Cognitively normal | Cognitively normal | Left ICA occlusion |
| Scanner | Biograph mMR (PET/MR) | Biograph mMR (PET/MR) | Biograph 40 (PET/CT) | ECAT HR+ 962 (PET) | Biograph Vision (PET/CT) |
| Image reconstruction | 3D OP-OSEM | 3D OP-OSEM | DIFT | FBP | 3D OP-OSEM |
| Voxel size | $1.4 \times 1.4 \times 2.0$ mm$^3$ | $1.4 \times 1.4 \times 2.0$ mm$^3$ | $2.3 \times 2.3 \times 2.0$ mm$^3$ | $2.0 \times 2.0 \times 2.4$ mm$^3$ | $1.8 \times 1.8 \times 1.8$ mm$^3$ |
| Number of subjects | 24 | 10 | 10 | 10 | 1 (16 noisy samples) |

and registered using PMOD software v4.3 and then checked by a radiological technologist.

The $^{15}$O-CO$_2$ data for observing the cerebral blood flow were obtained from the Akita Cerebrospinal and Cardiovascular Center in Japan [37]. This study was approved by the ethics committee of our research institution (reference number: 20–16). A patient with left internal carotid artery (ICA) occlusion aspirated at 1.5 GBq/min of $^{15}$O-CO$_2$ for 1 min on a Siemens Biograph Vision. The full-count image was reconstructed using a 3D OP-OSEM algorithm with time-of-flight information and point spread function modeling to $128 \times 128 \times 128$ voxels with a voxel size of $1.8 \times 1.8 \times 1.8$ mm$^3$ using all the list-mode data for the 3 min period and ultra-low dose PET image was obtained by randomly downsampling to 1/128th of the list-mode data. The corresponding T1-weighted MR images acquired individually were interpolated and registered using the normalized mutual information criteria of the SPM tool.

### B. Evaluation Metrics

We used only $^{18}$F-AV-45 data from 24 subjects for self-supervised pre-training. DIP-based PET image denoising of the main task was evaluated using $^{18}$F-AV-45, $^{11}$C-PIB, and $^{18}$F-FDG data from every 10 subjects, but not from the pre-training dataset. As the ultra-low dose $^{15}$O-CO$_2$ data were generated by downsampling the full list-mode data, we generated independent 16 noisy samples and evaluated their denoising performance.

To evaluate the denoising performance quantitatively, we calculated the contrast-to-noise ratio (CNR) improvement ratio (CNRIR) between the denoised and original PET images of $^{18}$F-AV-45, $^{11}$C-PIB, and $^{18}$F-FDG, which has been commonly used in previous studies without a reference image [17], [33].

$$\text{CNRIR} = \frac{\text{CNR}_d - \text{CNR}_o}{\text{CNR}_o} \times 100\%, \qquad (4)$$

where the $\text{CNR}_d$ and $\text{CNR}_o$ are CNR of the denoised and original PET images, respectively. CNR is defined as:

$$\text{CNR} = \frac{|\overline{S_g} - \overline{S_w}|}{\sqrt{\sigma_g^2 + \sigma_w^2}}, \qquad (5)$$

where $\overline{S_g}$, $\overline{S_w}$ and $\sigma_g$, $\sigma_w$ represent the mean uptake and standard deviation (STD) corresponding to the region of interest (ROI) in the gray and white matter, respectively. In addition, a Wilcoxon signed-rank test was performed on the

CNRIR to statistically compare the performances of the different denoising methods.

In the $^{15}$O-CO$_2$ study, we calculated the trade-off between the contrast recovery coefficient (CRC) and STD in the normal right brain area.

$$\text{CRC} = \frac{1}{R} \sum_{r=1}^{R} \left( \frac{\overline{\alpha_r}}{\overline{\beta_r}} - 1 \right) \Big/ \left( \frac{\overline{\alpha_{full}}}{\overline{\beta_{full}}} - 1 \right), \qquad (6)$$

$$\text{STD} = \frac{1}{R} \sum_{r=1}^{R} \frac{1}{\overline{\beta_r}} \sqrt{\frac{1}{K_\beta} \sum_{k=1}^{K_\beta} (\beta_{r,k} - \overline{\beta_r})^2}, \qquad (7)$$

where $R$ represents the number of samples, $\overline{\alpha_r}$, $\overline{\alpha_{full}}$ and $\overline{\beta_r}$, $\overline{\beta_{full}}$ represent the mean uptakes in the right gray- and white-matter ROIs of the denoised and full-count images, respectively. Here, 6 mm (gray matter) and 10 mm (white matter) diameter ROIs were set manually on the MR image and were calculated via superimposition on the co-registered PET image.

### IV. RESULTS

Fig. 2 shows axial and sagittal PET images processed using different denoising methods for $^{18}$F-AV-45, $^{11}$C-PIB, and $^{18}$F-FDG. In addition, Fig. 3 shows the results of the box plots and Wilcoxon signed-rank test of the CNRIR calculated from the denoised images. The PET images processed using DIP with the pre-training model exhibited superior denoising performance while preserving the edges compared to conventional IGF and original DIP from scratch. Comparing the difference between pre-training models for DIP, the proposed denoising method yielded the highest value for CNRIR, and a statistical significance was noted between the previous MR2PET model and the self-supervised pre-training model for $^{11}$C-PIB ($p < 0.05$) and $^{18}$F-FDG ($p < 0.05$), excluding $^{18}$F-AV-45 ($p = 0.0527$).

Figs. 4 and 5 show the three orthogonal slices and contrast-to-noise tradeoff curves processed using different denoising methods for $^{15}$O-CO$_2$. Even with ultra-low dose $^{15}$O-CO$_2$ images, the proposed method achieved superior denoising performance with a higher contrast recovery and lower STD than the other denoising methods. The line profiles passing through the left ICA occlusion area in the denoised $^{15}$O-CO$_2$ PET images (see Fig. 6). The proposed method was capable of accurately restoring specific regions not included in the pre-training dataset, such as the inhaling mask (yellow arrows) and occlusion area (white arrows) (Fig. 4).

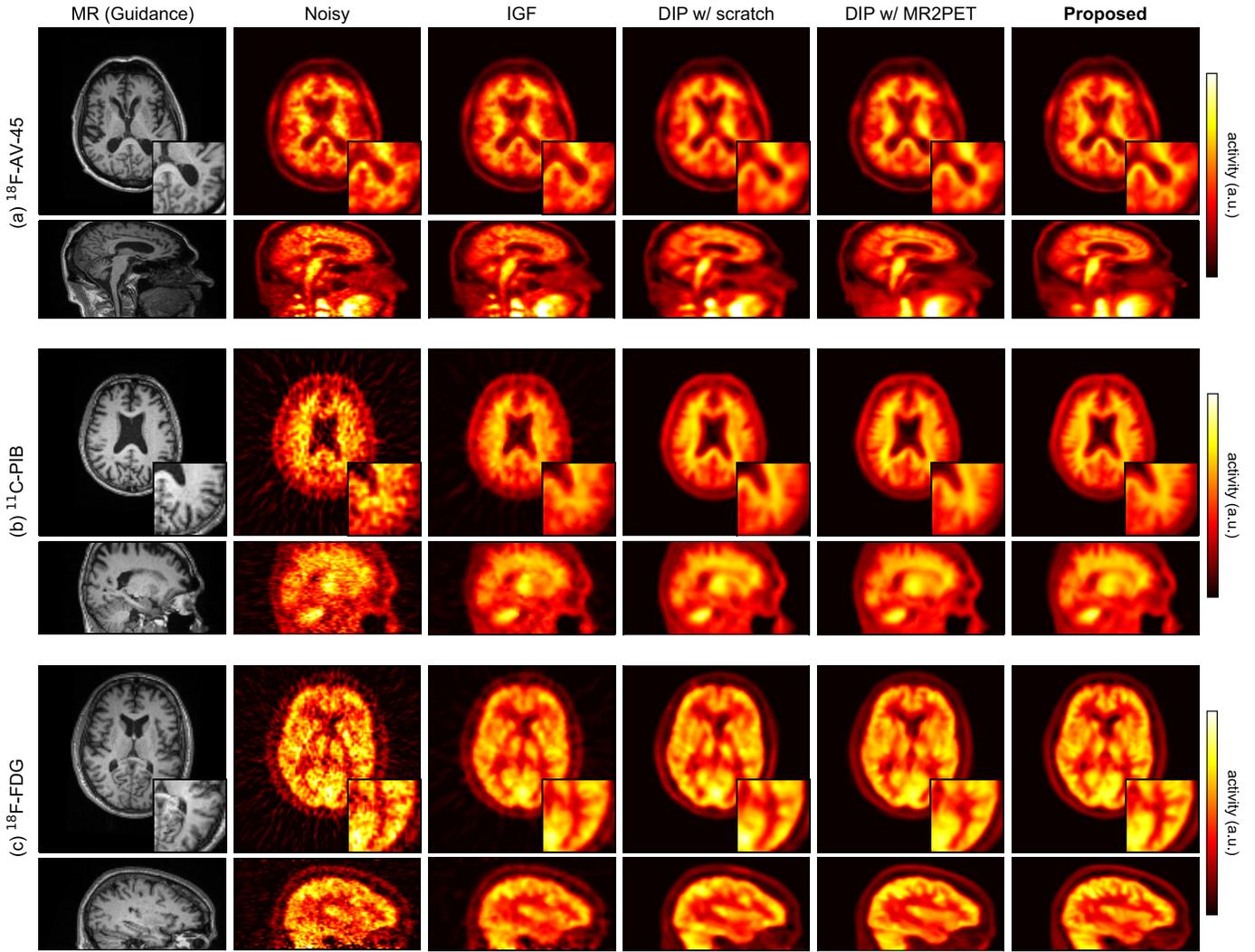

Fig. 2. Axial and sagittal sections from (a) $^{18}$F-AV-45, (b) $^{11}$C-PIB, and (c) $^{18}$F-FDG PET images processed by different denoising methods. From left to right, the sample images represent the T1-weighted MR, noisy, and denoised PET images corresponding to the IGF, DIP from scratch, DIP with MR2PET pre-training model, and proposed DIP with self-supervised pre-training models. The bottom right in axial sections shows the magnified images of the brain.

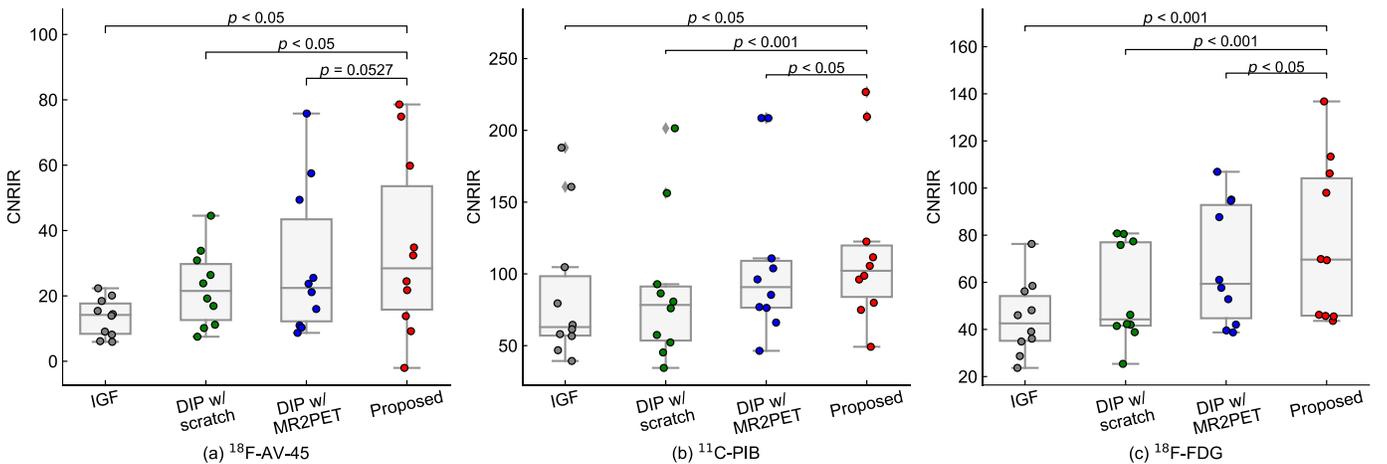

Fig. 3. Box plots and statistical analysis of CNRIRs calculated from the denoised (a) $^{18}$F-AV-45, (b) $^{11}$C-PIB, and (c) $^{18}$F-FDG images. The box shows the median and quartiles of data while the whiskers represent the rest of the distribution except for outliers.

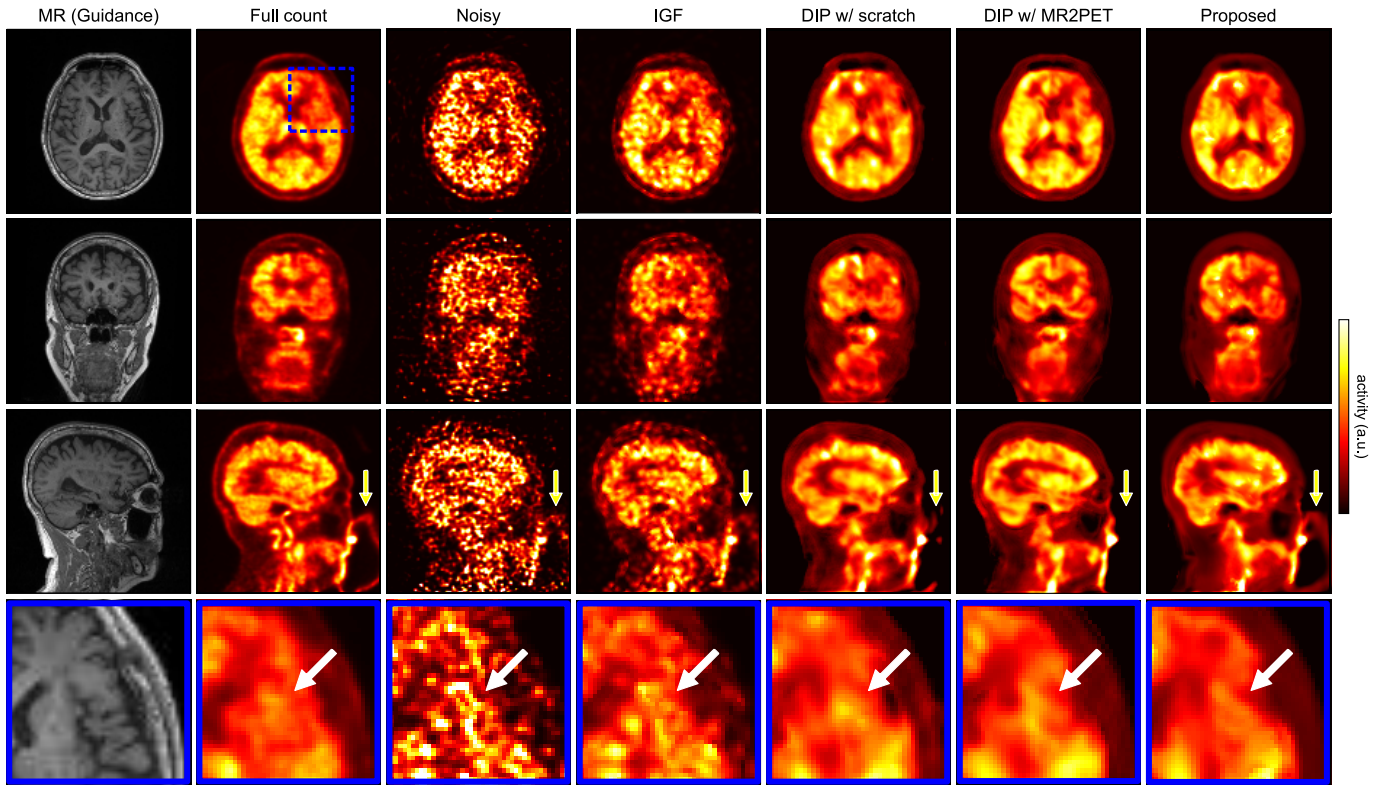

Fig. 4. Three orthogonal slices from $^{15}$O-CO$_2$ PET images processed by different denoising methods. From left to right, the sample images represent the T1-weighted MR, full count, noisy, and denoised PET images corresponding to the IGF, DIP from scratch, DIP with MR2PET pre-training model, and proposed DIP with self-supervised pre-training model. The bottom row shows the magnified images of the area marked by a blue rectangle. The proposed method showed an accurate restoration for specific regions such as inhaling mask (yellow arrows) and occlusion area (white arrows).

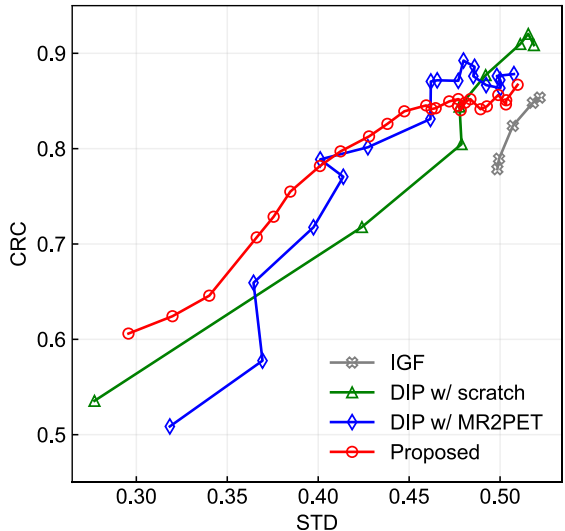

Fig. 5. CRC-STD tradeoff curves in the normal brain area of $^{15}$O-CO$_2$ PET images processed by different denoising methods. Markers are plotted corresponding to 1e-01, 1e-03, …, 1e-05 epsilons in the IGF, every epoch from 3 to 10 in the DIP w/ scratch, every epoch from 4 to 22 in the DIP w/ MR2PET, and every epoch from 5 to 30 in the proposed method.

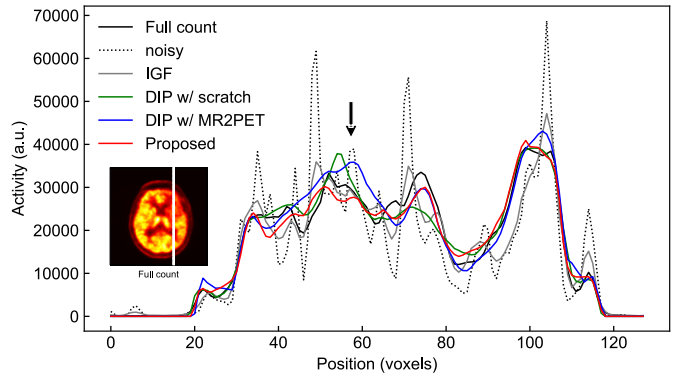

Fig. 6. Line profiles passing through the left occlusion area in the denoised $^{15}$O-CO$_2$ PET images by different denoising methods. The proposed method achieved an accurate restoration in occlusion area (black arrow).

## V. DISCUSSION

We proposed a self-supervised pre-training model for DIP-based unsupervised PET image denoising. The DIP framework transferred from the network parameters of the self-supervised pre-training model achieved state-of-the-art performance compared with other unsupervised denoising methods. Moreover, the proposed method robustly addressed different PET probes such as $^{18}$F-AV-45 and $^{11}$C-PIB, which observes amyloid-β; $^{18}$F-FDG, which indicates glucose metabolism; $^{15}$O-CO$_2$, which reveals cerebral blood flow, whereas the general supervised approach for PET image denoising requires massive training-label pairs and must be trained to handle specific tasks. The proposed method is considered to be particularly effective for short-half-life nuclides, such as $^{11}$C and $^{15}$O, and probes with few applicable cases, which are not realistic with a supervised framework. As noisy PET images with low counts could be restored to a quality close to that of the full-count image, it is expected that the proposed method reduces the PET scan time or radiotracer dose significantly without affecting the patients.

As with the introduction of image representation, e.g., ImageNet pre-trained CNNs have led to a significant expansion of the potential of image recognition [38], and the adaptation of a pre-training model to DIP was effective (Fig. 2). Although the original DIP might converge to a local minimum because it was trained using randomly initialized parameters, the pre-training model acquired better initial parameters, which promoted training stability. There was no significant difference between the results of the proposed and previous methods when the datasets of the pre-training and downstream tasks were the same [Fig. 3(a)]. On the other hand, the proposed method outperformed the previous method when the generality of both pre-training models was tested with PET probes that were not included in the pre-training dataset [Figs. 3(b) and 3(c)]. This indicates that the proposed self-supervised pre-training model was able to acquire transferable and more generalizable feature representations for PET images. The model learnt various spatial features such as context and texture information in PET images by spatially interpolating different degraded patterns in the pretext task. In particular, a non-linear transformation scheme has been reported to be able to learn the appearance of anatomical structures [28], [29] and, here, it is considered that the model acquired robust features for various PET images with different image contrasts.

According to the results shown (Figs. 4, 5, and 6), it was difficult to accurately restore the DIP from scratch at an ultra-low dose with a high noise level. Although the denoising performance was improved using pre-training models with basis representations of PET images, the previous method could not adapt to the inhaling mask and occlusion regions. Because the previous pre-training method aimed to acquire the basis functions contained in conditional DIP by domain transformation from MR to PET, downstream DIP is more sensitive to structural differences such as diseases between MR and PET images when qualitative variation in the pre-training dataset is poor. Different structures tend to be restored with a delay in the DIP optimization process, making it difficult to determine the optimal stopping criterion that matched the timing at which the noise is most removed. The proposed pre-training was performed only on PET images without MR images based on a 3D image restoration task. Therefore, the above challenges can be avoided. In addition, there is the advantage that it is not necessary to retroactively collect a large number of PET and MR images of the same patient for a pre-training dataset, enabling adopt easily accessible public datasets.

One limitation of this study was that we evaluated only the denoising aspects of DIP using a self-supervised pre-training model. In recent years, several applied studies for DIP-based PET image reconstruction have been reported [22], [24]–[27]. We believe that the proposed pre-training model could be easily adapted to these frameworks. Furthermore, this pre-training method has the potential to be applied not only to PET image denoising and reconstruction but to all other deep-learning tasks. Further, our future work will evaluate the effectiveness of various PET imaging applications using a self-supervised pre-training model.

## CONCLUSION

We proposed a self-supervised pre-training model to improve the DIP-based PET image denoising performance. Our proposed self-supervised pre-training model acquires generalizable and transferable visual representations from only unlabeled PET images by restoring various degraded PET images. We evaluated the proposed method using clinical brain PET data with various radioactive tracers, $^{18}$F-AV-45, $^{11}$C-PIB, $^{18}$F-FDG, and $^{15}$O-CO$_2$ acquired from different PET scanners. The proposed method using the self-supervised pre-training model achieved a robust and state-of-the-art denoising performance while retaining spatial details and quantification accuracy compared with other unsupervised methods and pre-training model. These results highlight the potential that the proposed method is particularly effective against rare diseases and probes and helps reduce the scan time or radiotracer dose without affecting the patients.

## ACKNOWLEDGMENT

The authors thank the members of the fifth research group at the Central Research Laboratory of Hamamatsu Photonics K. K. and the staff at the Department of Radiology and Nuclear Medicine, Akita Cerebrospinal and Cardiovascular Center.